\documentclass[%
 reprint,
superscriptaddress,
 amsmath,amssymb,
 aps,
 prb,
]{revtex4-1}

\usepackage{sidecap}
\sidecaptionvpos{figure}{c}

\usepackage{graphicx}
\usepackage{dcolumn}
\usepackage{bm}
\usepackage[
colorlinks=true,
linkcolor=blue,
citecolor=blue,
filecolor=blue,
urlcolor=blue]{hyperref}

\usepackage{amsmath,amsfonts,
	amsgen,amsbsy,amsopn,amstext,amssymb
}
\usepackage{epstopdf}
\usepackage{color}

\renewcommand{\i}{i}
\renewcommand{\(}{\left(}
\renewcommand{\)}{\right)}
\renewcommand{\Pr}{\mathrm{Pr}}
\newcommand{\e}{\mathrm{e}}

\newcommand{\Ze}{\bar{Z_{e}}}

\begin{document}

\preprint{}

\title{Quasi-perfect absorption by sub-wavelength acoustic panels \\ in transmission using accumulation of resonances due to slow sound}

\author{No\'e Jim\'enez}\email{noe.jimenez@univ-lemans.fr}
\author{Vicent Romero-Garc{\'i}a}
\author{Vincent Pagneux}
\author{Jean-Philippe Groby}
\affiliation{Laboratoire d'Acoustique de l'Universit{\'e} du Maine - CNRS UMR 6613, Le Mans, France}


\date{\today}

\begin{abstract}
We theoretically and experimentally report sub-wavelength resonant panels for low-frequency quasi-perfect sound absorption including transmission by using the accumulation of cavity resonances due to the slow sound phenomenon. The sub-wavelength panel is composed of periodic horizontal slits loaded by identical Helmholtz resonators (HRs). Due to the presence of the HRs, the propagation inside each slit is strongly dispersive, with near-zero phase velocity close to the resonance of the HRs. In this slow sound regime, the frequencies of the cavity modes inside the slit are down-shifted and the slit behaves as a subwavelength resonator. Moreover, due to strong dispersion, the cavity resonances accumulate at the limit of the bandgap below the resonance frequency of the HRs. Near this accumulation frequency, simultaneously symmetric and antisymmetric quasi-critical coupling can be achieved. In this way, using only monopolar resonators quasi-perfect absorption can be obtained in a material including transmission. 
\end{abstract}

\maketitle


\section{Introduction}\label{sec:intro}

Two main type of audible acoustic panels are desired in practical engineering applications: first, non-reflecting treatments and, second, zero-transmission materials. The first group requires that the reflection coefficient of a rigidly backed material vanishes. This is typically achieved by using porous or fibrous materials, that are mainly efficient in the inertial regime and for frequencies higher than the so-called quarter wavelength resonance of the backed layer, i.e., $f=c_0/4 L$, the sound speed in current porous materials being nearly similar to that in the air, $c_0$. On the other hand, zero-transmission is commonly obtained by using highly reflecting materials together with an interior frame made of a porous absorber. These structures are efficient for frequencies higher than the first Fabry-P\'erot resonance of the slab of equivalent material, i.e. $f=c/2 L$. Concerning low frequency sound, both groups of structures result in practical limitations due to the excessive thickness, $L$, and weight of the treatments.

Designing materials with both zero-reflection and zero-transmission simultaneously, i.e., perfect absorbers including transmission, is of special interest. Perfect absorption is of particular interest for many applications such as energy conversion \cite{law2005}, time reversal technology \cite{derode1995}, coherent perfect absorbers \cite{chong2010} or soundproofing \cite{mei2012} among others. In the case of rigidly backed materials, acoustic metamaterials are efficient solutions to design sound absorbing materials which can present simultaneously sub-wavelength dimensions and strong or perfect acoustic absorption. These include double porosity materials \cite{olny2003}, metaporous materials \cite{groby2011,lagarrigue2013,boutin2013,groby2015b}, dead-end porosity materials \cite{dupont2011,leclaire2015}, metamaterials composed by membrane-type resonators \cite{yang2008,mei2012,ma2014,romero2016}, Helmholtz resonators (HRs) \cite{romero2016,romero2016use,jimenezAPL2016}, and quarter-wavelength resonators (QWRs) \cite{leclaire2015,groby2015,groby2016,li2016}. These last types of metamaterials \cite{leclaire2015,groby2015,groby2016,jimenezAPL2016} make use of strong dispersion giving rise to slow-sound propagation inside the material. Using slow sound results in a decrease of the cavity resonance frequency and, hence, the structure thickness can be drastically reduced to the deep-subwavelength regime \cite{groby2016}. 

In transmission problems, where both reflection and transmission are possible, acoustic metamaterials for perfect absorption have also been presented recently using decorated membrane resonators based on degenerate resonances \cite{yang2015}. In this case, monopolar and dipolar resonators are used to critically couple the symmetric and antisymmetric problem respectively and, therefore, to get perfect absorption of the full problem with transmission \cite{piper2014}. Perfect absorption has also been observed in metamaterials composed only by monopolar resonators, e.g. by using two unsymmetrical HRs \cite{merkel2015}. In these structures, the transmission vanishes at the resonance of one HR, behaving effectively as a hard wall. By tuning a second HR the reflection problem can be critically coupled and, therefore, perfect absorption is obtained.

In this work, we present quasi-perfect absorption in a sub-wavelength metamaterial panel for transmission problems using identical monopolar resonators. The current design is based on the accumulation of resonances due to slow sound. The system works as follows: first, strong dispersion inside the slits is generated below the resonance frequency of the HR, while around the resonance frequency a bandgap is generated and transmission vanishes. Second, in the propagation band the cavity resonances of each slit are stretched in frequency and accumulate below the resonance frequency of the HR. Due to this accumulation of resonances, the absorption using only monopolar resonators can exceed 50\%. Then, by tuning the geometry of the system it can be almost critically coupled with the exterior medium, therefore, achieving quasi-perfect absorption. 

\begin{SCfigure*}[1][htbp]
	\centering
	\includegraphics[width=12.5cm]{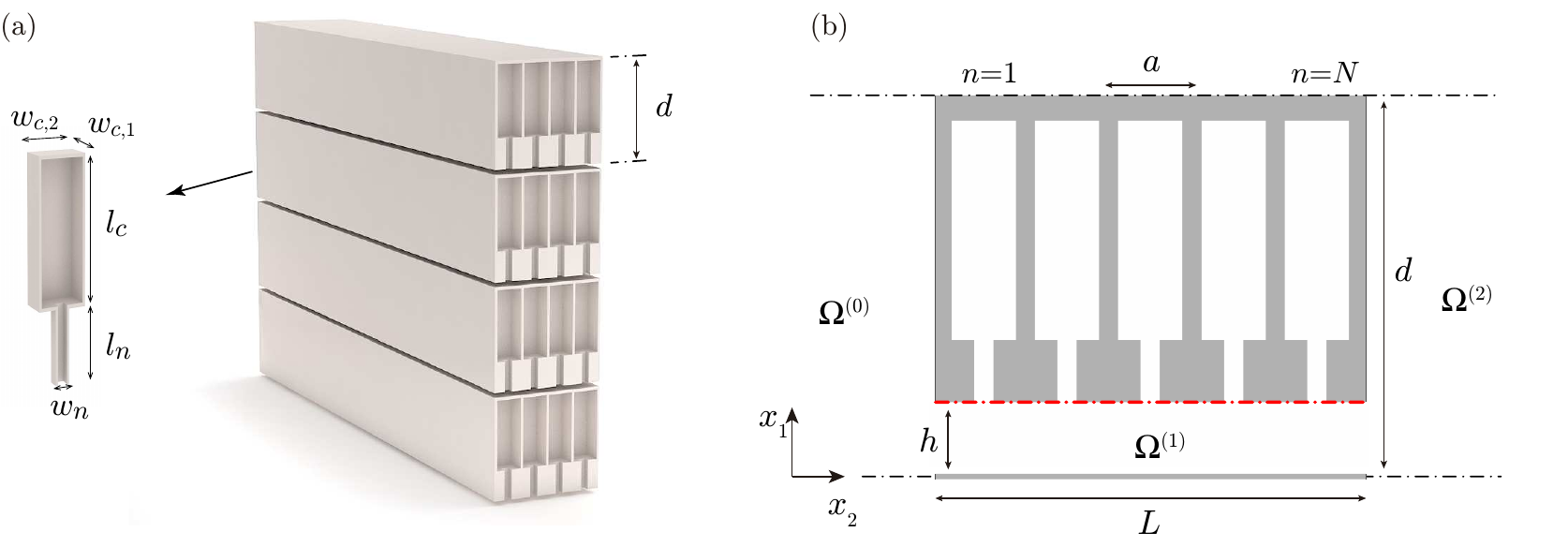}
	\caption{(a) Conceptual view of the thin panel placed on a rigid wall with $N=4$ layers of square cross-section Helmholtz resonators. (b) Scheme of the unit cell of the panel composed of a set of $N$ Helmholtz resonators. Symmetry boundary conditions are applied at boundaries $\Gamma_{x_1=d}$ and $\Gamma_{x_1=0}$.}
	\label{fig:scheme}
\end{SCfigure*}

\section{Description of the system and theoretical models}
The system consists in a panel perforated with a periodic arrangement of open slits, of thickness $h$ and length $L$, with periodicity $d$ in the $x_1$ direction, as shown in Fig.~\ref{fig:scheme}. The upper wall of the slit is loaded by $N$ identical HRs in an square array of side $a$, in such a way that the length of the system is $L=Na$. To maximize the volume of the cavity of the resonators an hence to minimize the resonance frequency, HRs with rectangular cross-section are used, characterized by a square neck with side $w_n$, a rectangular cavity with sides $w_{c,1}$ and $w_{c,2}$, and length $l_n$ and $l_c$ respectively. The visco-thermal losses in the system are considered in both the resonators and in the slit by using effective complex and frequency dependent physical parameters \cite{stinson1991}. Therefore, by changing the geometry, the intrinsic losses of the system can be tuned.

In order to have a deeper understanding of the physics involved in the system described above, several theoretical models with different hypothesis have been applied to analyze the scattering of the system. In this Section, we briefly present each one, more details being given in the appendix.

\subsection{Modal expansion method (MEM)}
The first model is a modal expansion method (MEM) \cite{groby2016}. The acoustic field is expanded in the modal basis for each domain, and then each domain is assembled by applying the boundary conditions (continuity of pressure and velocity are considered at the entrance and exit of the slit). The effect of the resonators is included by an impedance condition at $x_1=h$ in the slit, by considering $Z = Z_\mathrm{HR}/\phi$, with $Z_\mathrm{HR}$ the impedance of the HRs and $\phi=w_n^2/a^2$ the slit surface porosity. The impedance of the HR including the radiation correction of the neck is presented in the appendix. Eventually, a mode-matching linear system is obtained, allowing us to calculate the reflection and transmission coefficients \cite{groby2016}. 

One of the interesting point of this method is that considering the first order terms in the expansion, the low frequency approximation of the mode-matching system gives the effective parameters, i.e., the complex and frequency dependent effective bulk modulus, $\kappa_{e}$, and effective density, $\rho_{e}$, as:  
\begin{eqnarray}
\kappa_{e}&=&\frac{{\kappa_s}}{\phi_{t}}{\left[1+\frac{\kappa_s\phi\left(V_c\kappa_n+V_n\kappa_c\right)}{\kappa_n h\left(S_n\kappa_c-V_c\rho_n l_n \omega^2\right)}\right]}^{-1}, \label{eq:ke}\\ 
\rho_{e}&=&\frac{\rho_s}{\phi_{t}}, \, \label{eq:rhoe}
\end{eqnarray}
\noindent where $\phi_t=h/d$ is the total porosity of the metamaterial, $\rho_s$ and $\rho_n$ are the effective densities of the slit and neck, $\kappa_s$, $\kappa_n$ and $\kappa_c$ are the effective bulk modulus of the slit, neck and cavity respectively, $V_n$ and $V_c$ are the volumes of the neck and cavity of the HRs respectively and $\omega$ is the angular frequency. From Eqs.~(\ref{eq:ke}-\ref{eq:rhoe}) it is clear that this metamaterial allows only for negative compressibility for frequencies around the resonance frequency of the HRs. The reflection and transmission coefficients are linked to the effective parameters by
\begin{eqnarray}
R_t &=& \frac{i (\Ze^2-1)\sin (k_e L)}{2\Ze\cos (k_eL) - i(\Ze^2+1)\sin (k_e L)} \,, \\
T_t  &=& \frac{2 \Ze}{2\Ze\cos (k_eL) - i(\Ze^2+1)\sin (k_e L)}\,,
\end{eqnarray}
\noindent with normalized effective impedance $\Ze=\rho_{e}\kappa_{e}/\rho_{0}\kappa_{0}$ and effective wavenumber $k_{e}=\omega\sqrt{\rho_{e}/\kappa_{e}}$. Finally, the absorption of the system is defined as $\alpha = 1-|T_t|^2-|R_t|^2$. Another interesting point of the MEM is that if high order terms are included in the expansion, no end corrections are required both at the entrance and exit of the slit.

\subsection{Transfer matrix method (TMM)}
The second model is based on the transfer matrix method (TMM), in which the $N$ resonators are included as 1D point scatterers. $N$ matrices are assembled for a system made of $N$ unit cells, giving the transfer matrix, from which the reflection, transmission and absorption coefficients can be obtained. Using the transfer matrix of a single unit cell with the Floquet-Bloch periodic boundary conditions, the dispersion relations in the periodic system can be obtained. This model accounts for the finite and discrete arrangement of HR while the MEM model, being based on an average impedance at the wall, does not consider the effect of discreteness, i.e., the finite number of HR. Moreover, contrary to the MEM, in which we can obtain the effect of the higher order modes in the system, the TMM directly considers the plane wave approximation, which is valid for the low frequency regime. The radiation corrections are included in the impedances of the resonators and in the slits in order to mimic the effect of the higher order modes. Importantly, the TMM gives a direct information of the effect of the finite number of resonators. 

\subsection{Finite Element Method (FEM)}
In order to validate the previous analytical models and to show the limits of validity of each of them, we have used a finite element method (FEM) algorithm ({COMSOL Multiphysics\color{red}}) to solve the scattering problem, where the effects of the finite number of resonators, the higher order modes and the losses are taken into account numerically. The visco-thermal losses are introduced as effective parameters in the slit and the resonator elements \cite{stinson1991}. A plane wave impinges the system and the complete geometry is considered using the radiation conditions that simulates the Sommerfeld conditions in the limits of the numerical domains.

\section{Dispersion relation and slow sound conditions in the slit}

\begin{figure}[b]
	\centering
	\includegraphics[width=8.7cm]{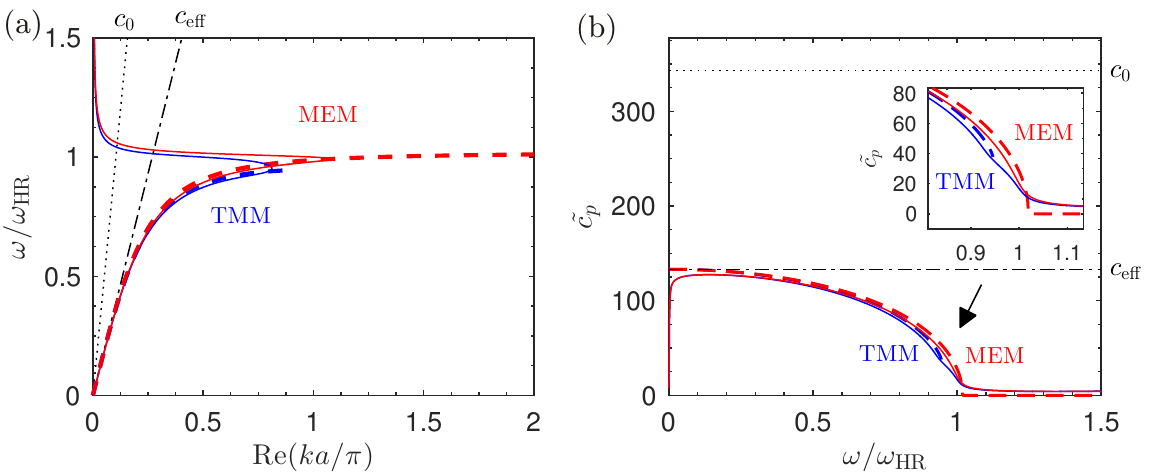}
	\caption{(a) Dispersion relation in the slit calculated using (red) modal expansion method (MEM), (blue) transfer matrix method (TMM) in the lossless (dashed line) and lossy case (continuous line). Wavenumber in the air $k_0=\omega/c_0$ (dotted line) and the low frequency asymptotic limit $\omega/c_\mathrm{eff}$ (dashed-dotted line). (b) Representation of $\tilde{c_p}=\mathrm{Re}(\omega/k)$, which is closely related to the phase speed. The inset shows the slow sound regime.}
	\label{fig:ceff}
\end{figure}

Figure \ref{fig:ceff}~(a) shows the real part of the complex wavenumber in the slit using MEM and TMM for a metamaterial with parameters $h=1.2$ mm, $a=4.95$ cm, $w_n=7.1$ mm, $w_c=4.9$, $d=5$ cm, $l_n=7.3$ mm, and $l_c=d-h-l_n$. A band-gap is observed above the resonance frequency of the HRs, $\omega_\mathrm{HR}$. Inside the slit a dispersive propagation band is generated below the band-gap frequency, where the wavenumber in the slit is remarkably increased and, therefore, as Fig.~\ref{fig:ceff}~(b) shows, slow sound conditions are achieved in this range of frequencies. It can be observed that the effective low-frequency sound speed in the lossless case, calculated from Eqs.(\ref{eq:ke}-\ref{eq:rhoe}), $c_\mathrm{eff}={c_0}/{\sqrt{1+ {V_\mathrm{tot}\phi_n}/{h S_n}}}$, is accurately described by both models, with $V_\mathrm{tot}$ the total volume of the resonator. 

\begin{figure*}[t]
	\centering
	\includegraphics[width=\textwidth]{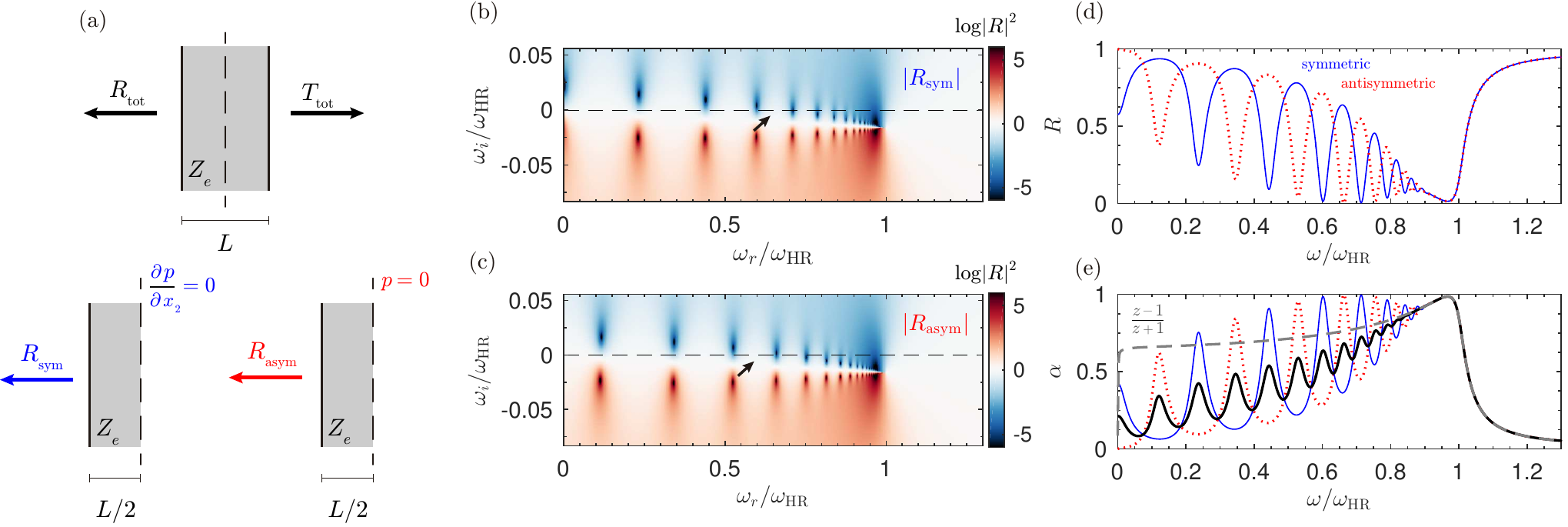}
	\caption{(a) Symmetric and antisymmetric problem decomposition for an homogeneous layer of material with effective parameters. (b-c) Complex frequency representation od the reflection coefficient for the symmetric and antisymmetric problem respectively. (d) Reflection coefficient at the real axis for symmetric (blue) and antisymmetric (dotted red). (e) Absorption for symmetric (blue), antisymmetric (dotted red), total (thick black) and impedance matching condition (dashed gray).}
	\label{fig:Sym}
\end{figure*}

In the lossless case, zero phase velocity can be observed for frequencies just below $\omega_\mathrm{HR}$. It is worth noting here that the maximum wavenumber using TMM is limited to $k_\mathrm{max}=\pi/a$ due to the fact that this model accounts for the periodicity of the system, while using MEM the wavenumber can be infinite in the lossless case, due to the fact that the MEM does not account for the periodicity of the system. This limitation of the MEM will be discussed in more detail later in Sec. \ref{sec:discrete}.

Once thermoviscous losses are introduced, the dispersion relation inside the slit is modified. As Theocharis et al. showed \cite{theocharis2014}, the minimum value of speed of sound is limited by the thermoviscous losses. Figure~\ref{fig:ceff}~(b) shows the slow sound region below $\omega_\mathrm{HR}$. We can clearly see that, although the quantity $\tilde{c}_p=\mathrm{Re}(\omega/k)$, closely related to the phase velocity, is anymore zero in the lossy case, slow sound conditions are still observed in this range of frequencies: the mean phase speed in the low frequency range is much lower than the speed of sound in air, $c_0$, and the phase speed near the resonance frequency of the HR is extremely low. This makes possible the shortening of the ratio $\lambda/L$ and, therefore, the slit behaves as a sub-wavelength resonator. The dispersion relation in the slit and the sound speed predicted by the different theoretical models agree in the low frequency regime, while only small differences can be observed near the band-gap between the TMM and MEM calculations due to the different hypothesis used in each model.

\section{Reflection and transmission problems}

In this work we deal with a symmetric and reciprocal system. In the low frequency regime the problem can be considered as 1D because only plane waves propagate. In this situation, the two eigenvalues of the scattering matrix (with $T_t$ on the diagonal), which relates the amplitudes of the input with the output waves, are given by $T_t-R_t$ and $T_t+R_t$. Therefore the absorption of the system can be described by decomposing the full problem in its symmetric and antisymmetric equivalent problems in reflection \cite{merkel2015}, as shown schematically in Fig.~\ref{fig:Sym}~(a). In fact, the two eigenvalues of the scattering matrix give the reflection coefficients for the symmetric and antisymmetric sub-problems in reflection, respectively. Thus, by setting rigid, ${\partial p}/{\partial x}=0$ (symmetric), and soft, $p=0$ (anti-symmetric), boundary conditions at the symmetry plane of the system, the reflection coefficients of each sub-problem can be obtained as
\begin{eqnarray}
R_\mathrm{sym}&=&T_t-R_t=\frac{\Ze \sin k_e L/2-i \cos k_e L/2}{\Ze\sin k_e L/2 + i \cos k_e L/2}, \\
R_\mathrm{asym}&=&T_t+R_t=\frac{\Ze \cos k_e L/2 + i \sin k_e L/2}{i\sin k_e L/2 - \Ze\cos k_e L/2} ,
\end{eqnarray}
\noindent for the symmetric and antisymmetric problems respectively. Then, the absorption coefficient of the full problem can be obtained from the absorption of each subproblem as, $\alpha=(\alpha_\mathrm{sym}+\alpha_\mathrm{asym})/2$, where $\alpha_\mathrm{sym(asym)}=1-|R_\mathrm{sym(asym)}|^2$ \cite{merkel2015}.

\begin{figure*}[t]
	\centering
	\includegraphics[width=18cm]{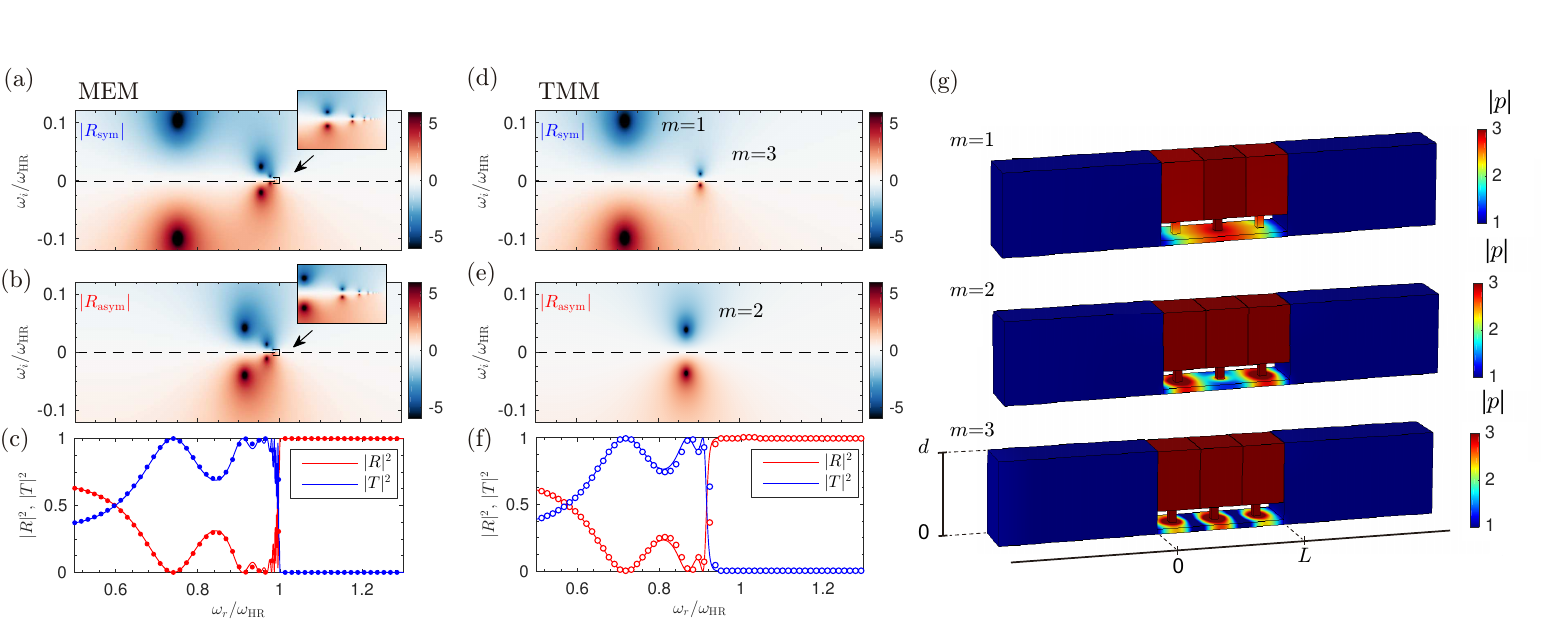}
	\caption{Complex frequency plane representation of the reflection coefficient for a panel of $N=3$ resonators. (a) Symmetric and (b) asymmetric reflection obtained using the modal expansion method (MEM). Colorbar in $\log |R|^2$ units. (c) Total transmission (blue) and reflection (red) using MEM (continuous lines) and its effective parameters (dots). (d) Symmetric and (e) asymmetric reflection using the transfer matrix method (TMM). (f) Total transmission (blue) and reflection (red) using TMM. (g) Acoustic field obtained using finite element method (FEM) at frequencies corresponding to the resonances $m=1,2,3$, colorbar in normalized pressure units.}
	\label{fig:Models}
\end{figure*}

\subsection{Asymptotic behaviour, large number of resonators}
\label{sec:assymtotic}

Let us first consider $N$ sufficiently large to accurately describe the system as a slab of material with the effective parameters. Figures~\ref{fig:Sym}~(b-c) show the corresponding reflection coefficient in the complex frequency plane. It is obtained with the TMM, of the symmetric and antisymmetric problems for $N=30$ resonators considering a complex frequency $\omega=\omega_r + i \omega_i$, with $\omega_r$ and $\omega_i$ the real and imaginary frequencies. First, it can be observed that a series of zero-pole pairs appear in the frequency complex plane \cite{romero2016use}. The poles correspond to the cavity modes inside the slab of effective material \cite{pagneux2013}. Due to dispersion, these cavity modes accumulate below the resonance frequency of the HRs. On the other hand, it can be seen that the cavity modes of the symmetric problem (see $R_\mathrm{sym}$) appear at frequencies different from the frequencies of the antisymmetric one (see $R_\mathrm{asym}$). This effect is clearly seen in Figure~\ref{fig:Sym}~(d), where the reflection coefficients for each problem are plotted at the real axis of frequencies.

In addition, it can be seen that for some particular frequencies, as those marked with the arrows in Fig.~\ref{fig:Sym}~(b-c), the zeros of the reflection coefficient are located on the real axis of frequencies. At these frequencies, the reflection coefficient of the (anti)symmetric vanishes and the structure is critically coupled. This condition is enough to achieve perfect absorption for, e.g., the symmetric problem, as was demonstrated in rigid-backed materials \cite{romero2016,groby2016,jimenezAPL2016}. However, for obtain perfect absorption of the full problem including transmission, both symmetric and antisymmetric reflection coefficients must simultaneously vanish \cite{piper2014,merkel2015}, as the following relations hold: $R_t=(R_\mathrm{sym}+R_\mathrm{asym})/{2}$ and $T_t=(R_\mathrm{sym}-R_\mathrm{asym})/{2}$. 

In general, for a homogeneous slab of material the cavity resonances of the symmetric and antisymmetric problems, i.e., its Fabry-P\'erot modes, are staggered in frequency and perfect absorption is not possible. However, in our system the cavity modes are accumulated below the limit of the band-gap because of the strong dispersion introduced by the presence of the resonators. Then, the zeros of the reflection coefficient for the symmetric and antisymmetric problems can be close one to another in frequency and quasi-perfect absorption can be obtained at the edge of the band-gap. Figure~\ref{fig:Sym}~(e) shows the corresponding absorption of the full problem (black line), where the absorption due to accumulation of resonances around $\omega_\mathrm{HR}$ is observed. It is interesting to show that, in the limit of a semi-infinite panel, both the reflection coefficient of the symmetric and antisymmetric problems collapse to the impedance matching condition, $\lim_{L\to\infty}R_\mathrm{sym}=\lim_{L\to\infty}R_\mathrm{asym}={(\Ze-1)}/{(\Ze +1)}
$, and then, only in this limit, perfect absorption can be achieved, as shown in Figure~\ref{fig:Sym}~(e). However, for a finite layer $R_\mathrm{sym} \ne R_\mathrm{asym}$ and only quasi-perfect absorption can be reached with a single homogenized slab of material. Moreover, $\Ze$ is generally complex and no perfect matching can be achieved.

\subsection{Finite number of resonators}\label{sec:discrete}
\begin{SCfigure*}[1][htbp]
	\centering
	\includegraphics[width=13cm]{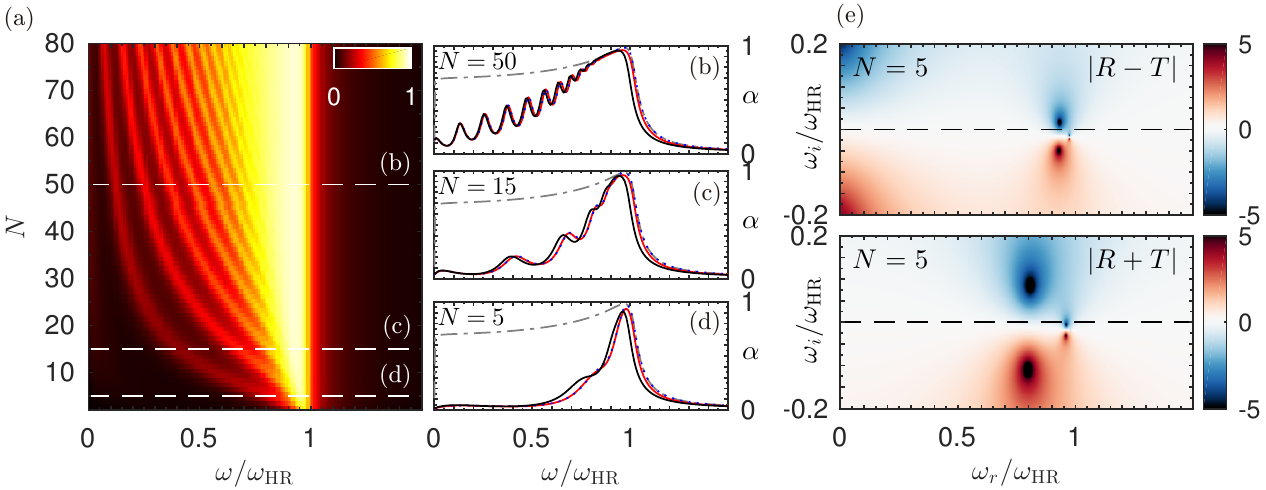}
	\caption{(a) Absorption of the material as a function of the number of resonators and frequency. (b-d) Absorption for $N=50,15$ and $5$ resonators, obtained using MEM (red), effective parameters (dotted blue) and TMM (black). Dashed-dotted gray line shows the impedance matching condition. (e) Complex frequency representation of the reflection coefficient for $N=5$ resonators. Colormap in $\log |R|^2$ units.}
	\label{fig:Nmap}
\end{SCfigure*}

Figure~\ref{fig:Models} shows the scattering of the system in the lossless case for $N=3$ resonators with the same parameters as in the Section \ref{sec:assymtotic}. Figure~\ref{fig:Models}~(a-b) shows the complex frequency representation of the reflection coefficient obtained by using the MEM, while Fig.~\ref{fig:Models}~(d-e) shows the corresponding one obtained by using the TMM. It is worth noting here that if we solve the problem using the MEM (with an effective-impedance boundary condition to represent the effect of the HRs), it presents a zero-pole structure with an infinite collection of resonances that accumulate around the band-gap frequency. Note for a finite number of resonators there should be a finite number of resonances. The TMM correctly accounts for the finite number of resonances, in this case $N=3$, in agreement with FEM simulations as shown in Fig.~\ref{fig:Models}~(g). It is also worth noting here that these cavity resonances are in fact the collective modes of the HRs and there exist only $N$ different collective modes. Fig.~\ref{fig:Models}~(f) shows the total reflection and transmission in the real axis calculated with TMM and FEM. The agreement between the two methods is very good, showing that the hypothesis of the TMM are correct for the considered frequency range. It can also be noticed that, by including the discreteness of the system, the band-gap frequency is shifted down from $\omega_\mathrm{HR}$, as it was previously noticed from the dispersion relations. However, the finite number of HRs limits the accumulation of resonances near the band-gap: as $N$ decreases the condition to have symmetric and antisymmetric resonances close to one another in frequency will become more difficult to achieve. Therefore, the number of (identical) HRs is a critical parameter to obtain quasi-perfect absorption in metamaterials with transmission by means of the accumulation of resonances.

Once losses are introduced in the system, the zero-pole structure is down shifted in the complex frequency plane and the system starts to absorb energy \cite{romero2016use}. Figure~\ref{fig:Nmap}~(a) presents the absorption of a panel as a function of the number of resonators, $N$, and frequency. First, it can be observed that for a relatively large number of resonators quasi-perfect absorption can be achieved even when the discreteness is retained, e.g. for $N=50$ resonators as shown in the Fig.~\ref{fig:Nmap}~(b). The material is almost impedance matched with the exterior medium and both MEM and TMM agree: the system can be described as an homogenized slab of locally reactive material. Only small differences between MEM and TMM solutions exist near the band-gap due to the infinite number of resonances of the MEM.

For most sound absorption applications it is desirable to use of panels with reduced thickness, and of special interest is the design of panels with subwavelength dimensions. Then, when reducing the panel thickness the number of resonators must also be reduced and, therefore, the accumulation of resonances becomes limited. Figure~\ref{fig:Nmap}~(c) shows the absorption of a panel with $N=15$ ($L\approx \lambda_{\alpha_\mathrm{max}}/2$), while Fig.~\ref{fig:Nmap}~(d) shows the absorption of a panel with $N=5$ ($L=\lambda_{\alpha_\mathrm{max}}/6.6$). In both cases, a peak of absorption is still observed, but its amplitude falls to $\alpha_\mathrm{max}=0.96$ and $\alpha_\mathrm{max}=0.92$ respectively for each case. The corresponding reflection coefficient in complex frequency plane for $N=5$ is shown in Fig.~\ref{fig:Nmap}~(f) for the symmetric and antisymmetric problems. Again, the zeros of the symmetric problem appear staggered in frequency with respect to the antisymmetric one and, as a consequence, the accumulation of resonances is limited by the small number of resonators. However, by tuning the geometry of the system, high acoustic absorption can be achieved by locating one zero of the reflection coefficient of the symmetric problem on the real frequency axis and, simultaneously, locate another zero of the antisymmetric problem as close as possible to the real axis at a different but nearly frequency. Therefore, the maximum value of absorption is directly dependent on the number of HRs and inversely dependent on the panel thickness. Using an array of identical resonators the design of the panel is a compromise between the peak acoustic absorption and the parameter $\lambda_\mathrm{max}/L$.

\section{Experimental results}

\begin{SCfigure*}[1][tbp]
	\centering
	\includegraphics[width=12cm]{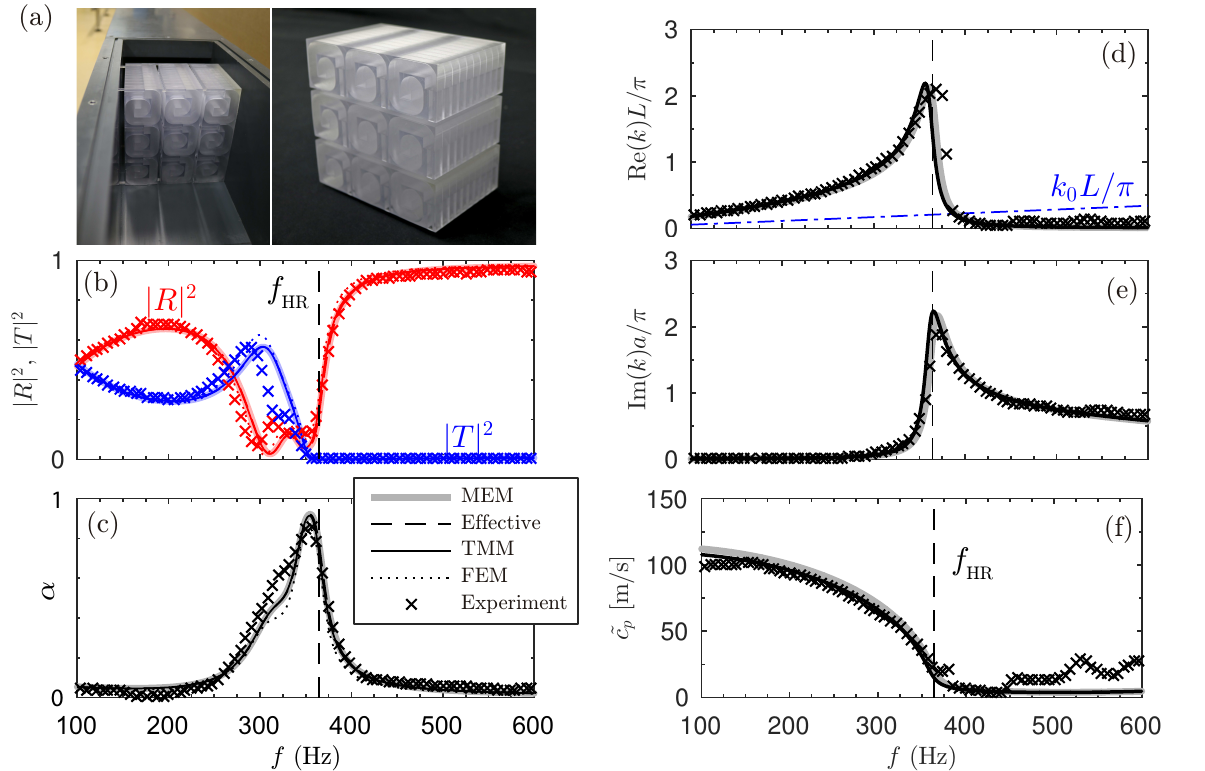}
	\caption{(a) Photographs of the experimental setup, where the semitransparent resin allows to see the coiled Helmholtz resonators inside the material. (b) Corresponding reflection (red) and transmission (blue) coefficients of the sample measured experimentally (markers), calculated using the MEM (thick lines), by using the effective parameters (dashed), TMM (continuous) and finite element simulation (dotted). The vertical dashed line marks the resonance of the HRs. (c) Corresponding absorption. (d) Real part of the real part of the wavenumber (thick gray), and its reconstruction using the experimental data (markers) and analytic data (continuous line). (e) Corresponding imaginary part. (f) Quantity $\tilde{c_p}=\mathrm{Re}(\omega/k)$, closely related to the speed of sound. }
	\label{fig:experiment}
\end{SCfigure*}

A subwavelength thickness sample with $N=10$ resonators was built using stereolithography techniques using a photosensitive epoxy polymer (Accura 60{\textsuperscript{\textregistered}}, 3D Systems Corporation, Rock Hill, SC 29730, USA), where the acoustic properties of the solid phase are $\rho_0=1210$ kg/m$^3$, $c_0= 1630 \pm 60$ m/s. The geometry of the structure was tuned using an optimization method (sequential quadratic programming (SQP) method \cite{powell1978}) in order to maximize the absorption at a given frequency (350 Hz), while the panel thickness was constrained to $L=\lambda/10$. The resulting parameters were $h=4.3$ mm, $a=9.8$ mm, $w_n=5.3$ mm, $w_{c,1}=11.4$ mm, $w_{c,2}=9.3$ mm, $d=$ cm, $l_n=25.2$ mm, and $l_c=139.6$ mm. It is worth noting here that we use the coiling of the HRs in order to save space. Figure~\ref{fig:experiment} summarizes the experimental results. First, Fig.~\ref{fig:experiment}~(a) shows a photograph of the panel, composed by 3 unit cells with $N=10$ for each one, allowing the measurement of reflection and transmission coefficients at normal incidence, which are shown in Fig.~\ref{fig:experiment}~(b). A good agreement between the experimental results, theoretical predictions and FEM simulations is observed. The results show the band-gap generated by the resonance of the HRs, where the low-cutoff frequency of the band-gap is just below the resonance frequency of the HRs, $f_{\mathrm{HR}}=364$ Hz. In this frequency range, transmission almost vanishes and the total reflection does not, as shown in Fig.~\ref{fig:experiment}~(b), as a consequence of the staggered structure of zero-pole structure. The corresponding absorption is plotted in Fig.~\ref{fig:experiment}~(c), where again good agreement can be observed between theory and experiments. Here, at 350 Hz the absorption peak obtained from the experiments was $\alpha=0.87$, while $\alpha=0.91$ was obtained from TMM predictions. In addition, small differences can be observed around 300 Hz. These small discrepancies can be associated to imperfections on the fitting of the structure to the impedance tube and due to the coiling of the HRs. 

On the other hand, the effective wavenumber inside the slits was reconstructed using an inversion method \cite{groby2010}. Figure~\ref{fig:experiment}~(d,e) shows the experimental and theoretical reconstruction of the real and imaginary part of the wavenumber respectively. It can be observed that the experimental reconstruction agrees with the theoretical prediction. Here, at $f=350$ Hz where the peak absorption is observed, the real part of the wavenumber is greatly increased compared to the wavenumber in air, $k_0$. Moreover, the imaginary part of the wavenumber is also increased, leading to the damping of the acoustic waves inside the material. Finally, the quantity $\tilde{c}_p=\mathrm{Re}(\omega/k)$, is shown in Fig.~\ref{fig:experiment}~(f). It can be seen that slow sound conditions are achieved by the experiment and the speed of sound inside the material is reduced to $\tilde{c}_p=34$ m/s at the peak absorption frequency, $f=350$ Hz.

\section{Conclusions}

The absorption of panels in transmission with periodic arrays of slits loaded by monopolar resonant inclusions made of HRs has been studied. We have shown that by using an array of identical monopolar resonators the symmetric and antisymmetric resonances of the system exist at staggered frequencies and, therefore, simultaneous critical coupling of the symmetric and antisymmetric modes is not possible. However, due to the loading HRs, strong dispersion is observed in the interior of each slit and the cavity resonances accumulate below the band-gap frequency, being the symmetrical and antisymmetrical modes staggered but very close in frequency. In this frequency range and by tuning the geometry, the system can be quasi-critically coupled with the exterior medium and therefore quasi-perfect absorption can be obtained. 

The limits of acoustic absorption in symmetric and reciprocal panels with transmission were explored experimentally, and quasi-perfect absorption for a subwavelength panel with thickness $L=\lambda/10$ was demonstrated. These results underline the necessity of breaking the symmetry of the system to achieve perfect absorption in transmission or the use of degenerate resonators with symmetric (monopolar) and antisymmetric (dipolar) resonances. Strategies for breaking the symmetry include the use of, e.g, different sized HRs \cite{merkel2015}. Strategies to use degenerate resonances have been also studied \cite{piper2014,yang2015}.

On the other hand, the extension of the results shown in this work to other analogous physical systems is also possible, with special interest for those where the number of resonators in the cavity can be remarkably increased, e.g. by using deep-subwavelength resonators as arrays of identical membranes with added mass or by using gas micro-bubbles for underwater acoustic applications.

The proposed configuration presents interesting and remarkable features as the open slits allows the air to flow through the panel, e.g., being able to use these subwavelength metamaterials for low-frequency noise control in industrial applications where simultaneously chromatic noise absorption and machinery refrigeration or air flow are required. These promising results open the possibilities to study different configurations based on these metamaterials and to extend the results to broadband and omnidirectional perfect absorption with deep sub-wavelength structures, which remains a great scientific challenge. 

\begin{acknowledgments}	
This work has been funded by the Metaudible project ANR-13-BS09-0003, co-funded by ANR and FRAE.

\end{acknowledgments} 

\appendix

\section{Visco-thermal losses model}
The visco-thermal losses in the system are considered both in the resonators and in the slit by using its effective complex and frequency dependent parameters \cite{stinson1991}.

{\bfseries{Slits}}: The effective parameters in the slit, considering only plane waves propagate inside, are expressed as:
\begin{equation}\label{eq:rhos}
\rho_s={\rho _0}\left[1-\frac{\tanh \left(\frac{h}{2}{G_\rho}\right)}{\frac{h}{2}{G_\rho}}\right]^{-1} \,,
\end{equation}
\begin{equation}\label{eq:Ks}
\kappa_s=\kappa_0\left[1+(\gamma-1)\frac{\tanh \left(\frac{h}{2}{G_\kappa}\right) }{\frac{h}{2}{G_\kappa}}\right]^{-1} \,,
\end{equation}
\noindent with $G_\rho=\sqrt{{\i\omega\rho_0}/{\eta}}$ and $G_\kappa=\sqrt{\i\omega\mathrm{Pr}\rho_0/{\eta}}$, and where $\gamma$ is the specific heat ratio of air, $P_0$ is the atmospheric pressure, $\Pr$ is the Prandtl number, $\eta$ the dynamic viscosity, $\rho_0$ the air density and $\kappa_0={\gamma P_0}$ the air bulk modulus.

{\bfseries{Ducts}}: The propagation in a rectangular cross-section tube can be described by its complex and frequency dependent density and bulk modulus, and considering that plane waves propagate inside, can be expressed as \cite{stinson1991}:
\begin{equation}\label{eq:rhoc}
\rho_t = -\frac{\rho_0 a^2 b^2}{4 G_\rho^2 \sum\limits_{k\in\mathbb{N}}\sum\limits_{m\in\mathbb{N}} \left[\alpha_k^2 \beta_m^2 \(\alpha_k^2 + \beta_m^2 - G_\rho^2\) \right]^{-1}} \,,
\end{equation}
\begin{equation}\label{eq:Kc}
\kappa_t=\frac{\kappa_0}{\gamma + \frac{4 (\gamma -1) G_\kappa^2 }{{a^2}{b^2}}\sum\limits_{k\in\mathbb{N}}\sum\limits_{m\in\mathbb{N}}{\left[\alpha _k^2\beta _m^2\(\alpha _k^2+\beta _m^2-G_\kappa^2\)\right]^{-1}}} \,,
\end{equation}
\noindent with the constants $\alpha _k=2(k+1/2)\pi/a$ and $\beta_m=2(m+1/2)\pi/b$, and the dimensions of the duct $a$ and $b$ being either the neck, $a=b=w_n$, or the cavity, $a=w_{c,1}$ and $b=w_{c,2}$ of the Helmholtz resonators.

\section{Modal expansion method (MEM)}
\subsection{Mode matching system}\label{Sec:effective}

As sketched in Fig.\ref{fig:scheme}, the full space is divided into three sub-domains, the forward exterior medium, $\Omega^{[0]}$, the interior of the slit, $\Omega^{[1]}$, and the backward exterior air, $\Omega^{[2]}$. The field can be represented on each $\Omega^{[i]}$ domain as
\begin{eqnarray}\label{eq:fields1}
p^{[0]} &=& \sum_q \(A^i\delta _q \e^{-\i k_{2q}^{[0]}(x_2-L)} + R_q \e^{\i k_{2q}^{[0]}(x_2-L)}\)\e^{\i k_{1q} x_1} \,, \\
p^{[1]} &=& \sum_n \(B_n \e^{-\i k_{2n}^{[1]}x_2} + C_n \e^{\i k_{2n}^{[1]}x_2}\)\sin k_{1n}(x_1-L) \,, \label{eq:fields2}\\
p^{[2]} &=& \sum_q T_q \e^{\i k_{1q}x_1 - \i k_{2q}^{[0]}x_2} \label{eq:fields3}
\end{eqnarray}

\noindent where $A^i$ is the amplitude of the incident wave, $R_q$ and $T_q$ are the reflection and transmission coefficients of the $q$-th Bloch mode respectively and $\delta_n^0$ the Kronecker delta. Here and beyond the superscript ${[i]}$ indicates the domain according to Fig.~\ref{fig:scheme}. Periodic boundary conditions are assumed at $x_1=0$ and $x_1=d$. It is worth noting here that for normal incidence periodic boundary condition reduce to symmetric (rigid) boundary conditions. Inside the slit, at $x_1=h$, the effect of the resonators is included by a wall impedance condition given by $Z_\mathrm{wall}=Z_\mathrm{HR}/\phi$, with $Z_\mathrm{HR}$ the impedance of the HRs and $\phi=w_n^2/a^2$ the surface porosity of the slit. The application of these boundary conditions and continuity between domains leads to the following mode matching system:
\begin{eqnarray}\label{eq:Rqfull}
R_q &&- \sum_{q'}\sum_{n} \frac{\eta_n^{[1]}}{\eta_{nq}^{[0]}}\frac{h}{d N_n}\(\frac{-\i R_{q'}}{\tan k_{2n}^{[1]}L}\) I_{nq'}^+ I_{nq}^- 
\nonumber \\&&- \sum_{q''}\frac{\eta_n^{[1]}}{\eta_q^{[0]}}\frac{h}{d N_n} \sum_n\(\frac{\i T_{q''}}{\sin k_{2n}^{[1]}L}\)I_{nq''}^+I_{nq}^- = \nonumber \\
&&A^i + \sum_n\frac{\eta_n^{[1]}}{\eta_q^{[0]}}\frac{h}{d}\frac{-\i A^i}{\tan k_{2n}^{[1]}L N_n}I_{n0}^+I_{nq}^- \,,
\end{eqnarray}
\noindent and
\begin{eqnarray}\label{eq:Tqfull}
T_q &&+ \sum_{q'}\sum_n \frac{\eta_n^{[1]}}{\eta_{nq}^{[0]}}\frac{h}{d}\(\frac{-\i R_q}{N_n \sin k_{2n}^{[1]}L}\)I_{nq'}^+I_{nq}^- 
\nonumber\\&&+ \sum_{q''}\sum_n\frac{\eta_n^{[1]}}{\eta_{nq}^{[0]}}\frac{h}{d N_n}\frac{\i T_{q''}}{\tan k_{2n}^{[1]}L}I_{nq''}^+I_{nq}^- = \nonumber\\
&&\sum_n\frac{\eta_n^{[1]}}{\eta_{nq}^{[0]}}\frac{h}{d}\frac{\i A^i}{\sin (k_{2n}^{[1]}L)N_n}I_{n0}^+I_{nq}^- \,.
\end{eqnarray}

\noindent where $\eta_q^{[j]} = {k_{2q}^{[j]}}/{\rho^{[j]}}$. Thus, the reflection and transmission coefficients can be calculated by solving the linear system of Eqs.(\ref{eq:Rqfull}-\ref{eq:Tqfull}), where the integrals $I_{nq}^\pm$ are written in analogy with Refs. \cite{groby2015,groby2016,jimenezAPL2016}.

\subsection{Low frequency approximation: Effective parameters}

In the low frequency regime, the system (\ref{eq:Rqfull}-\ref{eq:Tqfull}) leads to the reflection and transmission coefficients as
\begin{eqnarray}\label{eq:memlow1}
R_0 + \frac{\i \Ze R_0}{\tan k_2^{[1]}L} - \frac{\i \Ze T_0}{\sin k_2^{[0]}L} &=& \(1 - \frac{\i \Ze}{\tan k_2^{[1]}L}\) \,,\\
T_0 - \frac{\i \Ze R_0}{\sin k_2^{[1]}L} + \frac{\i \Ze T_0}{\tan k_2^{[1]}L} &=& \frac{\i \Ze}{\sin k_2^{[1]}L} \,,\label{eq:memlow3}
\end{eqnarray}

\noindent where the normalized effective impedance is defined as $\Ze= {\eta^{[1]} h}/{\eta^{[0]} d} ={Z^{[1]}}/{Z^{[0]}\phi_t}$, 
with the total porosity $\phi_t={h}/{d}$, and $A^i=1$. We clearly identify the effective wavenumber inside the panel as
\begin{eqnarray}\label{eq:trans}
k_2^{[1]}=\sqrt{\(k_1^{[1]}\)^2 - \(k_{10}\)^2} \,.
\end{eqnarray}

The transversal component of the wavenumber is given by \cite{redon2011}
\begin{equation}
k_{10}=\frac{1}{h}\sqrt{\frac{-\i \omega \rho_s h}{Z_\mathrm{wall}}} \,,
\end{equation}

\noindent and using the Helmholtz resonator impedance in the low frequency regime it becomes
\begin{equation}
k_{10}=\frac{1}{h}\sqrt{\frac{-\omega\rho_s h\(V_c Z_n k_c + S_n Z_c k_n l_n\)\phi}{Z_n \(S_n Z_c - V_c Z_n k_c k_n l_n\)}}\,.
\end{equation}

Using Eq.~(\ref{eq:trans}) the effective wavenumber reads:
\begin{equation}
k_{e}=k_2^{[1]} = \frac{\omega^2}{c_0^2}\left[ 1 + \frac{\kappa_s\(V_c\kappa_n+V_n\kappa_c\)\phi}{\kappa_n\(S_n\kappa_c-V_cl_n\omega^2\rho_n\)h}\right] \,.
\end{equation}

The effective parameters, i.e. the complex and frequency dependent bulk modulus and density are given by
\begin{eqnarray}
\kappa_{e}&=&\frac{{\kappa_s}}{\phi_{t}}{\left[1+\frac{\kappa_s\phi\left(V_c\kappa_n+V_n\kappa_c\right)}{\kappa_n h\left(S_n\kappa_c-V_c\rho_n l_n \omega^2\right)}\right]}^{-1}, \\ 
\rho_{e}&=&\frac{\rho_s}{\phi_{t}}.
\end{eqnarray}

Using these parameters, the reflection and transmission coefficients can be calculated according to Eqs.~(\ref{eq:memlow1}-\ref{eq:memlow3}) as
\begin{eqnarray}
R_\mathrm{0} &=& \frac{i (\Ze^2-1)\sin (k_e L)}{2\Ze\cos (k_eL) - i(\Ze^2+1)\sin (k_e L)} \,, \\
T_\mathrm{0}  &=& \frac{2 \Ze}{2\Ze\cos (k_eL) - i(\Ze^2+1)\sin (k_e L)}\,.
\end{eqnarray}

\section{Transfer Matrix Method}
A discrete model is developed accounting for the finite number of resonators using the Transfer Matrix Method (TMM). Thus, for identical resonators, the transfer matrix is written as:
\begin{eqnarray}
\left( \begin{array}{c} 
{{P}_{{\rm i}}} \\ {{U}_{{\rm i}}} \end{array}\right)  
&=&\bf{T}
\left( \begin{array}{c} 
{{P}_{{\rm o}}} \\ 
{{U}_{{\rm o}}} 
\end{array} \right)\,,
\end{eqnarray}
\noindent where the transmission matrix $\bf{T}$ is written as
\begin{eqnarray}\label{eq:totalmatrix}
{\bf{T}}=\left( \begin{array}{cc} 
{{T}_{11}} & {{T}_{12}} \\ 
{{T}_{21}} & {{T}_{22}}
\end{array} \right) 
=  {\bf M}_{\Delta l_{\mathrm{slit}}}({\bf M}_{s}{\bf M}_\mathrm{HR}{\bf M}_{s})^{N}{\bf M}_{\Delta l_{\mathrm{slit}}}\,.\nonumber
\end{eqnarray}

\noindent Here, the transmission matrix for each lattice step in the slit, ${\bf M}_s$, is written as \renewcommand{\arraystretch}{2}
\begin{equation}
{{\bf M}_{s}}=\left( 
\begin{array}{cc} 
\cos\(k_s \dfrac{a}{2}\) & \i Z_s \sin \(k_s \dfrac{a}{2}\) \\ 
\dfrac{\i}{Z_s}\sin \(k_s \dfrac{a}{2}\) & \cos \(k_s \dfrac{a}{2}\) 
\end{array} \right),
\end{equation}
\noindent where the slit characteristic impedance is written as $Z_s=\sqrt{\kappa_s \rho_s}/S_s$ and $S_s=h\,a$. The resonators are introduced as a punctual scatters by a transmission matrix ${{\bf M}_{\mathrm{HR}}}$ as\renewcommand{\arraystretch}{1}
\begin{eqnarray}
{{\bf M}_{\mathrm{HR}}}=
\left( \begin{array}{cc} 
1 & 0 \\ 
1/{{Z}_{\mathrm{HR}}} & 1 
\end{array} \right),
\end{eqnarray}
\noindent and the radiation correction of the slit to the free space as 
\begin{eqnarray}
{\bf M}_{\Delta l_{\mathrm{slit}}}=
\left( \begin{array}{cc} 
1 & {Z}_{\Delta l_{\mathrm{slit}}} \\ 
0 & 1 
\end{array} \right),
\end{eqnarray}
\noindent with the characteristic radiation impedance of the slit ${Z}_{\Delta l_{\mathrm{slit}}}=-i\omega\Delta l_\mathrm{slit}\rho_0/\phi_t S_0$, where $S_0=d\,a$, $\rho_0$ the air density and $\Delta l_\mathrm{slit}$ the proper end correction that will be described later. 

The reflection and transmission coefficients of the system can be directly calculated from the elements of the matrix $\bf{T}$ as
\begin{eqnarray}
\label{eq:T}
T_\mathrm{TMM} &=& \frac{2e^{\imath k L}}{T_{11}+T_{12}/Z_0+Z_0T_{21}+T_{22}},\\
\label{eq:Rm}
R_\mathrm{TMM} &=& \frac{T_{11}+T_{12}/Z_0-Z_0T_{21}-T_{22}}{T_{11}+T_{12}/Z_0+Z_0T_{21}+T_{22}},
\end{eqnarray}
\noindent with $Z_0=\rho_0 c_0/S_0$, and the effective parameters can be obtained from the transfer matrix elements as follows
\begin{eqnarray}
k_{\mathrm{TMM}}&=&\frac{1}{L}\cos^{-1}\left(\frac{{T_{11}+T_{22}}}{2}\right)\,, \\
Z_{\mathrm{TMM}}&=&\frac{1}{Z_0}\sqrt{\frac{T_{12}}{T_{21}}}.
\end{eqnarray}

\section{Resonator impedance and End corrections}

Using the effective parameters for the neck and cavity elements given by Eqs.~(\ref{eq:rhoc}-\ref{eq:Kc}), The impedance of a Helmholtz resonator can be written as 
\begin{equation}
Z_\mathrm{HR}=i Z_n \frac{A-\tan k_n l_n \tan k_c l_c}{A \tan k_n l_n + \tan k_c l_c},
\end{equation}
\noindent with $A=Z_c /Z_n $, $l_n$ and $l_c$ are the neck and cavity lengths, $S_n=w_n^2$ and $S_c=w_{c,1}w_{c,2}$ are the neck and cavity surfaces and $k_n$ and $k_c$, and $Z_n$ and $Z_c$ are the effective wavenumbers and effective characteristic impedance in the neck and cavity respectively.

It is worth noting here that this expression is not exact as long as correction due to the radiation should be included. The characteristic impedance accounting for the neck radiation can be expressed as \cite{theocharis2014}:
\begin{widetext}
	\begin{equation}
	Z_\mathrm{HR} = -\i \frac{\cos(k_n l_n) \cos(k_c l_c) - Z_n k_n \Delta l \cos(k_n l_n) \sin(k_c l_c)/Z_c - Z_n \sin(k_n l_n)\sin(k_c l_c)/Z_c}{\sin(k_n l_n)\cos(k_c l_c)/Z_n - k_n \Delta l\sin(k_n l_n)\sin(k_c l_c)/Z_c + \cos(k_n l_n)\sin(k_c l_c)/Z_c}\,,
	\end{equation}
\end{widetext}
\noindent where the correction length is deduced from the addition of two correction lengths $\Delta l=\Delta l_1 + \Delta l_2$ as
\begin{eqnarray}
\Delta l_1 &=& 0.82 \left[1 - 1.35 \frac{r_n}{r_c} + 0.31 \(\frac{r_n}{r_c}\)^3\right] r_n \,,\\
\Delta l_2 &=& 0.82 \left[1 - 0.235 \frac{r_n}{r_s} - 1.32\(\frac{r_n}{r_t}\)^2 \right. \\
&&\left. + 1.54 \(\frac{r_n}{r_t}\)^3 - 0.86\(\frac{r_n}{r_t}\)^4 \right] r_n \,.
\end{eqnarray}

The first length correction, $\Delta l_1$, is due to pressure radiation at the discontinuity from the neck duct to the cavity of the Helmholtz resonator \cite{kergomard1987}, while the second $\Delta l_2$ comes from the radiation at the discontinuity from the neck to the principal waveguide \cite{dubos1999}. This correction only depends on the radius of the waveguides, so it becomes important when the duct length is comparable to the radius, i.e., for small neck lengths and for frequencies where $k r_n \ll 1$.

Another important end correction comes from the radiation from the slits to the free air. The radiation correction for a periodic distribution of slits can be expressed as \cite{mechel2013}: 
\begin{equation}\label{eq:Dslit}
\Delta l_{\rm slit} = h \phi_t \sum_{n=1}^{\infty}\frac{\sin^2\(n\pi \phi_t\)}{(n\pi \phi_t)^3}.
\end{equation}

\noindent Note for $0.1\le \phi_t \le 0.7$ this expression reduces to $\Delta l_{\rm slit}\approx-{\sqrt{2}}\ln\left[ \sin\({\pi \phi_t}/{2}\)\right]/{\pi}$. Although Eq.~(\ref{eq:Dslit}) is appropriate for a periodic array of slits, it is not exact for slits loading HRs, therefore, we can evaluate a more realistic value for the end correction by reconstructing an equivalent impedance, $\tilde{Z}$, from the reflection coefficient of the zeroth order Bloch mode calculated with the MEM and comparing it to $\Ze=\rho_{e}\kappa_{e}/\rho_{0}\kappa_{0}$ in analogy with Ref.\cite{groby2016}:
\begin{equation}\label{eq:Dslit2}
\tilde{Z} - \Ze = -i\omega\frac{\rho_0}{\phi_t} \Delta l_\mathrm{slit}
\end{equation}

The slit end correction using this last approach gives a value that depends on the geometry of the HRs and for the present structures is around 1.5 times the one using Eq.(\ref{eq:Dslit}).

\section{Reconstruction of effective parameters}
\begin{figure}[t]
	\centering
	\includegraphics[width=8.5cm]{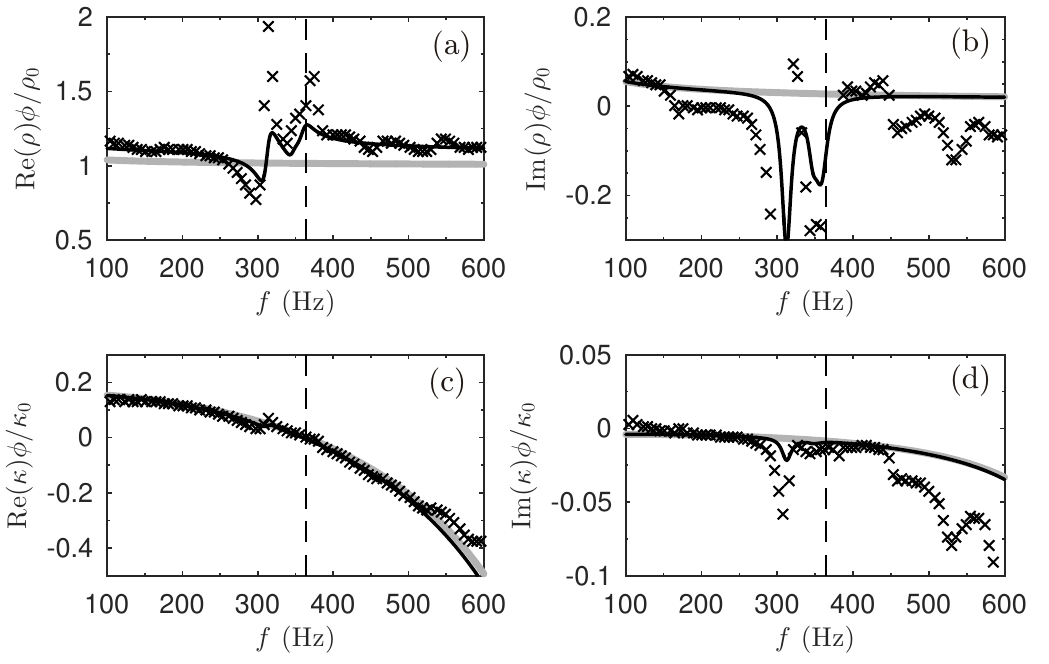}
	\caption{Effective parameters computed by (thick gray lines) the low frequency approximation of the MEM, Eqs.(\ref{eq:ke}-\ref{eq:rhoe}), (markers) reconstructed from the experimental data and (continuous lines) reconstructed from the MEM analytical reflection and transmission coefficients. (a) Real and (b) imaginary part of the complex density normalized by the ambient density and total porosity. (c) Real and (d) imaginary part of the complex compressibility normalized by the ambient bulk modulus and total porosity. The dashed line mark the resonant frequency of the Helmholtz resonators, $f_\mathrm{HR}=364$.}
	\label{fig:effective}
\end{figure}

The effective parameters of the metamaterial, i.e., the complex and frequency dependent density and bulk modulus, can be reconstructed from its effective wavenumber and impedance. Figure~\ref{fig:effective} shows the reconstruction using the experimental data (markers), the effective parameters given by Eqs.(\ref{eq:ke}-\ref{eq:rhoe}) (gray line) and the reconstruction method using the MEM analytical reflection and transmission coefficients as the input (black line). It can be seen that the inversion method, even with the analytical transmission and reflection data, fails to reconstruct the effective parameters at the frequencies corresponding to $\mathrm{Re}(k)L/\pi=(1,2,\ldots)$. This is caused by the poor reconstruction of the effective impedance at the Fabry-Perot resonances of a slab of effective material. However, the main features of the effective parameters are captured by the reconstruction. First, the effective density of the metamaterial is almost the density of air normalized by the total porosity, $\phi_t$, being it almost constant in frequency. Second, the real part of the effective bulk modulus is reduced up to 15 \% the bulk modulus of the air at 350 Hz, it vanish at the resonance frequency of the HRs ($f_\mathrm{HR}=364$ Hz, vertical dashed line), an it is negative for frequencies above $f_\mathrm{HR}$. Therefore, the effective compressibility of the material is greatly increased, allowing the structure to resonate in the subwavelength regime.

\bibliography{HRtransmission}

\begin{thebibliography}{31}%
\makeatletter
\providecommand \@ifxundefined [1]{%
 \@ifx{#1\undefined}
}%
\providecommand \@ifnum [1]{%
 \ifnum #1\expandafter \@firstoftwo
 \else \expandafter \@secondoftwo
 \fi
}%
\providecommand \@ifx [1]{%
 \ifx #1\expandafter \@firstoftwo
 \else \expandafter \@secondoftwo
 \fi
}%
\providecommand \natexlab [1]{#1}%
\providecommand \enquote  [1]{``#1''}%
\providecommand \bibnamefont  [1]{#1}%
\providecommand \bibfnamefont [1]{#1}%
\providecommand \citenamefont [1]{#1}%
\providecommand \href@noop [0]{\@secondoftwo}%
\providecommand \href [0]{\begingroup \@sanitize@url \@href}%
\providecommand \@href[1]{\@@startlink{#1}\@@href}%
\providecommand \@@href[1]{\endgroup#1\@@endlink}%
\providecommand \@sanitize@url [0]{\catcode `\\12\catcode `\$12\catcode
  `\&12\catcode `\#12\catcode `\^12\catcode `\_12\catcode `\%12\relax}%
\providecommand \@@startlink[1]{}%
\providecommand \@@endlink[0]{}%
\providecommand \url  [0]{\begingroup\@sanitize@url \@url }%
\providecommand \@url [1]{\endgroup\@href {#1}{\urlprefix }}%
\providecommand \urlprefix  [0]{URL }%
\providecommand \Eprint [0]{\href }%
\providecommand \doibase [0]{http://dx.doi.org/}%
\providecommand \selectlanguage [0]{\@gobble}%
\providecommand \bibinfo  [0]{\@secondoftwo}%
\providecommand \bibfield  [0]{\@secondoftwo}%
\providecommand \translation [1]{[#1]}%
\providecommand \BibitemOpen [0]{}%
\providecommand \bibitemStop [0]{}%
\providecommand \bibitemNoStop [0]{.\EOS\space}%
\providecommand \EOS [0]{\spacefactor3000\relax}%
\providecommand \BibitemShut  [1]{\csname bibitem#1\endcsname}%
\let\auto@bib@innerbib\@empty
\bibitem [{\citenamefont {Law}\ \emph {et~al.}(2005)\citenamefont {Law},
  \citenamefont {Greene}, \citenamefont {Johnson}, \citenamefont {Saykally},\
  and\ \citenamefont {Yang}}]{law2005}%
  \BibitemOpen
  \bibfield  {author} {\bibinfo {author} {\bibfnamefont {M.}~\bibnamefont
  {Law}}, \bibinfo {author} {\bibfnamefont {L.~E.}\ \bibnamefont {Greene}},
  \bibinfo {author} {\bibfnamefont {J.~C.}\ \bibnamefont {Johnson}}, \bibinfo
  {author} {\bibfnamefont {R.}~\bibnamefont {Saykally}}, \ and\ \bibinfo
  {author} {\bibfnamefont {P.}~\bibnamefont {Yang}},\ }\href@noop {} {\bibfield
   {journal} {\bibinfo  {journal} {Nat. Mater.}\ }\textbf {\bibinfo {volume}
  {4}},\ \bibinfo {pages} {455} (\bibinfo {year} {2005})}\BibitemShut {NoStop}%
\bibitem [{\citenamefont {Derode}\ \emph {et~al.}(1995)\citenamefont {Derode},
  \citenamefont {Roux},\ and\ \citenamefont {Fink}}]{derode1995}%
  \BibitemOpen
  \bibfield  {author} {\bibinfo {author} {\bibfnamefont {A.}~\bibnamefont
  {Derode}}, \bibinfo {author} {\bibfnamefont {P.}~\bibnamefont {Roux}}, \ and\
  \bibinfo {author} {\bibfnamefont {M.}~\bibnamefont {Fink}},\ }\href@noop {}
  {\bibfield  {journal} {\bibinfo  {journal} {Phys. Rev. Lett.}\ }\textbf
  {\bibinfo {volume} {75}},\ \bibinfo {pages} {4206} (\bibinfo {year}
  {1995})}\BibitemShut {NoStop}%
\bibitem [{\citenamefont {Chong}\ \emph {et~al.}(2010)\citenamefont {Chong},
  \citenamefont {Ge}, \citenamefont {Cao},\ and\ \citenamefont
  {Stone}}]{chong2010}%
  \BibitemOpen
  \bibfield  {author} {\bibinfo {author} {\bibfnamefont {Y.}~\bibnamefont
  {Chong}}, \bibinfo {author} {\bibfnamefont {L.}~\bibnamefont {Ge}}, \bibinfo
  {author} {\bibfnamefont {H.}~\bibnamefont {Cao}}, \ and\ \bibinfo {author}
  {\bibfnamefont {A.~D.}\ \bibnamefont {Stone}},\ }\href@noop {} {\bibfield
  {journal} {\bibinfo  {journal} {Phys. Rev. Lett.}\ }\textbf {\bibinfo
  {volume} {105}},\ \bibinfo {pages} {053901} (\bibinfo {year}
  {2010})}\BibitemShut {NoStop}%
\bibitem [{\citenamefont {Mei}\ \emph {et~al.}(2012)\citenamefont {Mei},
  \citenamefont {Ma}, \citenamefont {Yang}, \citenamefont {Yang}, \citenamefont
  {Wen},\ and\ \citenamefont {Sheng}}]{mei2012}%
  \BibitemOpen
  \bibfield  {author} {\bibinfo {author} {\bibfnamefont {J.}~\bibnamefont
  {Mei}}, \bibinfo {author} {\bibfnamefont {G.}~\bibnamefont {Ma}}, \bibinfo
  {author} {\bibfnamefont {M.}~\bibnamefont {Yang}}, \bibinfo {author}
  {\bibfnamefont {Z.}~\bibnamefont {Yang}}, \bibinfo {author} {\bibfnamefont
  {W.}~\bibnamefont {Wen}}, \ and\ \bibinfo {author} {\bibfnamefont
  {P.}~\bibnamefont {Sheng}},\ }\href@noop {} {\bibfield  {journal} {\bibinfo
  {journal} {Nat. Commun.}\ }\textbf {\bibinfo {volume} {3}},\ \bibinfo {pages}
  {756} (\bibinfo {year} {2012})}\BibitemShut {NoStop}%
\bibitem [{\citenamefont {Olny}\ and\ \citenamefont {Boutin}(2003)}]{olny2003}%
  \BibitemOpen
  \bibfield  {author} {\bibinfo {author} {\bibfnamefont {X.}~\bibnamefont
  {Olny}}\ and\ \bibinfo {author} {\bibfnamefont {C.}~\bibnamefont {Boutin}},\
  }\href@noop {} {\bibfield  {journal} {\bibinfo  {journal} {J. Acoust. Soc.
  Am.}\ }\textbf {\bibinfo {volume} {114}},\ \bibinfo {pages} {73} (\bibinfo
  {year} {2003})}\BibitemShut {NoStop}%
\bibitem [{\citenamefont {Groby}\ \emph {et~al.}(2011)\citenamefont {Groby},
  \citenamefont {Duclos}, \citenamefont {Dazel}, \citenamefont {Boeckx},\ and\
  \citenamefont {Lauriks}}]{groby2011}%
  \BibitemOpen
  \bibfield  {author} {\bibinfo {author} {\bibfnamefont {J.-P.}\ \bibnamefont
  {Groby}}, \bibinfo {author} {\bibfnamefont {A.}~\bibnamefont {Duclos}},
  \bibinfo {author} {\bibfnamefont {O.}~\bibnamefont {Dazel}}, \bibinfo
  {author} {\bibfnamefont {L.}~\bibnamefont {Boeckx}}, \ and\ \bibinfo {author}
  {\bibfnamefont {W.}~\bibnamefont {Lauriks}},\ }\href@noop {} {\bibfield
  {journal} {\bibinfo  {journal} {J. Acoust. Soc. Am.}\ }\textbf {\bibinfo
  {volume} {129}},\ \bibinfo {pages} {3035} (\bibinfo {year}
  {2011})}\BibitemShut {NoStop}%
\bibitem [{\citenamefont {Lagarrigue}\ \emph {et~al.}(2013)\citenamefont
  {Lagarrigue}, \citenamefont {Groby}, \citenamefont {Tournat}, \citenamefont
  {Dazel},\ and\ \citenamefont {Umnova}}]{lagarrigue2013}%
  \BibitemOpen
  \bibfield  {author} {\bibinfo {author} {\bibfnamefont {C.}~\bibnamefont
  {Lagarrigue}}, \bibinfo {author} {\bibfnamefont {J.}~\bibnamefont {Groby}},
  \bibinfo {author} {\bibfnamefont {V.}~\bibnamefont {Tournat}}, \bibinfo
  {author} {\bibfnamefont {O.}~\bibnamefont {Dazel}}, \ and\ \bibinfo {author}
  {\bibfnamefont {O.}~\bibnamefont {Umnova}},\ }\href@noop {} {\bibfield
  {journal} {\bibinfo  {journal} {J. Acoust. Soc. Am.}\ }\textbf {\bibinfo
  {volume} {134}},\ \bibinfo {pages} {4670} (\bibinfo {year}
  {2013})}\BibitemShut {NoStop}%
\bibitem [{\citenamefont {Boutin}(2013)}]{boutin2013}%
  \BibitemOpen
  \bibfield  {author} {\bibinfo {author} {\bibfnamefont {C.}~\bibnamefont
  {Boutin}},\ }\href@noop {} {\bibfield  {journal} {\bibinfo  {journal} {J.
  Acoust. Soc. Am.}\ }\textbf {\bibinfo {volume} {134}},\ \bibinfo {pages}
  {4717} (\bibinfo {year} {2013})}\BibitemShut {NoStop}%
\bibitem [{\citenamefont {Groby}\ \emph
  {et~al.}(2015{\natexlab{a}})\citenamefont {Groby}, \citenamefont
  {Lagarrigue}, \citenamefont {Brouard}, \citenamefont {Dazel}, \citenamefont
  {Tournat},\ and\ \citenamefont {Nennig}}]{groby2015b}%
  \BibitemOpen
  \bibfield  {author} {\bibinfo {author} {\bibfnamefont {J.-P.}\ \bibnamefont
  {Groby}}, \bibinfo {author} {\bibfnamefont {C.}~\bibnamefont {Lagarrigue}},
  \bibinfo {author} {\bibfnamefont {B.}~\bibnamefont {Brouard}}, \bibinfo
  {author} {\bibfnamefont {O.}~\bibnamefont {Dazel}}, \bibinfo {author}
  {\bibfnamefont {V.}~\bibnamefont {Tournat}}, \ and\ \bibinfo {author}
  {\bibfnamefont {B.}~\bibnamefont {Nennig}},\ }\href@noop {} {\bibfield
  {journal} {\bibinfo  {journal} {J. Acoust. Soc. Am.}\ }\textbf {\bibinfo
  {volume} {137}},\ \bibinfo {pages} {273} (\bibinfo {year}
  {2015}{\natexlab{a}})}\BibitemShut {NoStop}%
\bibitem [{\citenamefont {Dupont}\ \emph {et~al.}(2011)\citenamefont {Dupont},
  \citenamefont {Leclaire}, \citenamefont {Sicot}, \citenamefont {Gong},\ and\
  \citenamefont {Panneton}}]{dupont2011}%
  \BibitemOpen
  \bibfield  {author} {\bibinfo {author} {\bibfnamefont {T.}~\bibnamefont
  {Dupont}}, \bibinfo {author} {\bibfnamefont {P.}~\bibnamefont {Leclaire}},
  \bibinfo {author} {\bibfnamefont {O.}~\bibnamefont {Sicot}}, \bibinfo
  {author} {\bibfnamefont {X.~L.}\ \bibnamefont {Gong}}, \ and\ \bibinfo
  {author} {\bibfnamefont {R.}~\bibnamefont {Panneton}},\ }\href@noop {}
  {\bibfield  {journal} {\bibinfo  {journal} {J Appl Phys}\ }\textbf {\bibinfo
  {volume} {110}},\ \bibinfo {pages} {094903} (\bibinfo {year}
  {2011})}\BibitemShut {NoStop}%
\bibitem [{\citenamefont {Leclaire}\ \emph {et~al.}(2015)\citenamefont
  {Leclaire}, \citenamefont {Umnova}, \citenamefont {Dupont},\ and\
  \citenamefont {Panneton}}]{leclaire2015}%
  \BibitemOpen
  \bibfield  {author} {\bibinfo {author} {\bibfnamefont {P.}~\bibnamefont
  {Leclaire}}, \bibinfo {author} {\bibfnamefont {O.}~\bibnamefont {Umnova}},
  \bibinfo {author} {\bibfnamefont {T.}~\bibnamefont {Dupont}}, \ and\ \bibinfo
  {author} {\bibfnamefont {R.}~\bibnamefont {Panneton}},\ }\href@noop {}
  {\bibfield  {journal} {\bibinfo  {journal} {J. Acoust. Soc. Am.}\ }\textbf
  {\bibinfo {volume} {137}},\ \bibinfo {pages} {1772} (\bibinfo {year}
  {2015})}\BibitemShut {NoStop}%
\bibitem [{\citenamefont {Yang}\ \emph {et~al.}(2008)\citenamefont {Yang},
  \citenamefont {Mei}, \citenamefont {Yang}, \citenamefont {Chan},\ and\
  \citenamefont {Sheng}}]{yang2008}%
  \BibitemOpen
  \bibfield  {author} {\bibinfo {author} {\bibfnamefont {Z.}~\bibnamefont
  {Yang}}, \bibinfo {author} {\bibfnamefont {J.}~\bibnamefont {Mei}}, \bibinfo
  {author} {\bibfnamefont {M.}~\bibnamefont {Yang}}, \bibinfo {author}
  {\bibfnamefont {N.}~\bibnamefont {Chan}}, \ and\ \bibinfo {author}
  {\bibfnamefont {P.}~\bibnamefont {Sheng}},\ }\href@noop {} {\bibfield
  {journal} {\bibinfo  {journal} {Phys. Rev. Lett.}\ }\textbf {\bibinfo
  {volume} {101}},\ \bibinfo {pages} {204301} (\bibinfo {year}
  {2008})}\BibitemShut {NoStop}%
\bibitem [{\citenamefont {Ma}\ \emph {et~al.}(2014)\citenamefont {Ma},
  \citenamefont {Yang}, \citenamefont {Xiao}, \citenamefont {Yang},\ and\
  \citenamefont {Sheng}}]{ma2014}%
  \BibitemOpen
  \bibfield  {author} {\bibinfo {author} {\bibfnamefont {G.}~\bibnamefont
  {Ma}}, \bibinfo {author} {\bibfnamefont {M.}~\bibnamefont {Yang}}, \bibinfo
  {author} {\bibfnamefont {S.}~\bibnamefont {Xiao}}, \bibinfo {author}
  {\bibfnamefont {Z.}~\bibnamefont {Yang}}, \ and\ \bibinfo {author}
  {\bibfnamefont {P.}~\bibnamefont {Sheng}},\ }\href@noop {} {\bibfield
  {journal} {\bibinfo  {journal} {Nat. Mater.}\ }\textbf {\bibinfo {volume}
  {13}},\ \bibinfo {pages} {873} (\bibinfo {year} {2014})}\BibitemShut
  {NoStop}%
\bibitem [{\citenamefont {Romero-Garc{\'\i}a}\ \emph
  {et~al.}(2016{\natexlab{a}})\citenamefont {Romero-Garc{\'\i}a}, \citenamefont
  {Theocharis}, \citenamefont {Richoux}, \citenamefont {Merkel}, \citenamefont
  {Tournat},\ and\ \citenamefont {Pagneux}}]{romero2016}%
  \BibitemOpen
  \bibfield  {author} {\bibinfo {author} {\bibfnamefont {V.}~\bibnamefont
  {Romero-Garc{\'\i}a}}, \bibinfo {author} {\bibfnamefont {G.}~\bibnamefont
  {Theocharis}}, \bibinfo {author} {\bibfnamefont {O.}~\bibnamefont {Richoux}},
  \bibinfo {author} {\bibfnamefont {A.}~\bibnamefont {Merkel}}, \bibinfo
  {author} {\bibfnamefont {V.}~\bibnamefont {Tournat}}, \ and\ \bibinfo
  {author} {\bibfnamefont {V.}~\bibnamefont {Pagneux}},\ }\href@noop {}
  {\bibfield  {journal} {\bibinfo  {journal} {Sci. Rep.}\ }\textbf {\bibinfo
  {volume} {6}},\ \bibinfo {pages} {19519} (\bibinfo {year}
  {2016}{\natexlab{a}})}\BibitemShut {NoStop}%
\bibitem [{\citenamefont {Romero-Garc{\'\i}a}\ \emph
  {et~al.}(2016{\natexlab{b}})\citenamefont {Romero-Garc{\'\i}a}, \citenamefont
  {Theocharis}, \citenamefont {Richoux},\ and\ \citenamefont
  {Pagneux}}]{romero2016use}%
  \BibitemOpen
  \bibfield  {author} {\bibinfo {author} {\bibfnamefont {V.}~\bibnamefont
  {Romero-Garc{\'\i}a}}, \bibinfo {author} {\bibfnamefont {G.}~\bibnamefont
  {Theocharis}}, \bibinfo {author} {\bibfnamefont {O.}~\bibnamefont {Richoux}},
  \ and\ \bibinfo {author} {\bibfnamefont {V.}~\bibnamefont {Pagneux}},\
  }\href@noop {} {\bibfield  {journal} {\bibinfo  {journal} {The Journal of the
  Acoustical Society of America}\ }\textbf {\bibinfo {volume} {139}},\ \bibinfo
  {pages} {3395} (\bibinfo {year} {2016}{\natexlab{b}})}\BibitemShut {NoStop}%
\bibitem [{\citenamefont {Jim{\'e}nez}\ \emph {et~al.}(2016)\citenamefont
  {Jim{\'e}nez}, \citenamefont {Huang}, \citenamefont {Romero-Garc{\'\i}a},
  \citenamefont {Pagneux},\ and\ \citenamefont {Groby}}]{jimenezAPL2016}%
  \BibitemOpen
  \bibfield  {author} {\bibinfo {author} {\bibfnamefont {N.}~\bibnamefont
  {Jim{\'e}nez}}, \bibinfo {author} {\bibfnamefont {W.}~\bibnamefont {Huang}},
  \bibinfo {author} {\bibfnamefont {V.}~\bibnamefont {Romero-Garc{\'\i}a}},
  \bibinfo {author} {\bibfnamefont {V.}~\bibnamefont {Pagneux}}, \ and\
  \bibinfo {author} {\bibfnamefont {J.-P.}\ \bibnamefont {Groby}},\ }\href@noop
  {} {\bibfield  {journal} {\bibinfo  {journal} {Applied Physics Letters}\
  }\textbf {\bibinfo {volume} {109}},\ \bibinfo {pages} {121902} (\bibinfo
  {year} {2016})}\BibitemShut {NoStop}%
\bibitem [{\citenamefont {Groby}\ \emph
  {et~al.}(2015{\natexlab{b}})\citenamefont {Groby}, \citenamefont {Huang},
  \citenamefont {Lardeau},\ and\ \citenamefont {Aur{\'e}gan}}]{groby2015}%
  \BibitemOpen
  \bibfield  {author} {\bibinfo {author} {\bibfnamefont {J.-P.}\ \bibnamefont
  {Groby}}, \bibinfo {author} {\bibfnamefont {W.}~\bibnamefont {Huang}},
  \bibinfo {author} {\bibfnamefont {A.}~\bibnamefont {Lardeau}}, \ and\
  \bibinfo {author} {\bibfnamefont {Y.}~\bibnamefont {Aur{\'e}gan}},\
  }\href@noop {} {\bibfield  {journal} {\bibinfo  {journal} {J. Appl. Phys.}\
  }\textbf {\bibinfo {volume} {117}},\ \bibinfo {pages} {124903} (\bibinfo
  {year} {2015}{\natexlab{b}})}\BibitemShut {NoStop}%
\bibitem [{\citenamefont {Groby}\ \emph {et~al.}(2016)\citenamefont {Groby},
  \citenamefont {Pommier},\ and\ \citenamefont {Aur{\'e}gan}}]{groby2016}%
  \BibitemOpen
  \bibfield  {author} {\bibinfo {author} {\bibfnamefont {J.-P.}\ \bibnamefont
  {Groby}}, \bibinfo {author} {\bibfnamefont {R.}~\bibnamefont {Pommier}}, \
  and\ \bibinfo {author} {\bibfnamefont {Y.}~\bibnamefont {Aur{\'e}gan}},\
  }\href@noop {} {\bibfield  {journal} {\bibinfo  {journal} {J. Acoust. Soc.
  Am.}\ }\textbf {\bibinfo {volume} {139}},\ \bibinfo {pages} {1660} (\bibinfo
  {year} {2016})}\BibitemShut {NoStop}%
\bibitem [{\citenamefont {Li}\ and\ \citenamefont {Assouar}(2016)}]{li2016}%
  \BibitemOpen
  \bibfield  {author} {\bibinfo {author} {\bibfnamefont {Y.}~\bibnamefont
  {Li}}\ and\ \bibinfo {author} {\bibfnamefont {B.~M.}\ \bibnamefont
  {Assouar}},\ }\href@noop {} {\bibfield  {journal} {\bibinfo  {journal} {Appl.
  Phys. Lett.}\ }\textbf {\bibinfo {volume} {108}},\ \bibinfo {pages} {063502}
  (\bibinfo {year} {2016})}\BibitemShut {NoStop}%
\bibitem [{\citenamefont {Yang}\ \emph {et~al.}(2015)\citenamefont {Yang},
  \citenamefont {Meng}, \citenamefont {Fu}, \citenamefont {Li}, \citenamefont
  {Yang},\ and\ \citenamefont {Sheng}}]{yang2015}%
  \BibitemOpen
  \bibfield  {author} {\bibinfo {author} {\bibfnamefont {M.}~\bibnamefont
  {Yang}}, \bibinfo {author} {\bibfnamefont {C.}~\bibnamefont {Meng}}, \bibinfo
  {author} {\bibfnamefont {C.}~\bibnamefont {Fu}}, \bibinfo {author}
  {\bibfnamefont {Y.}~\bibnamefont {Li}}, \bibinfo {author} {\bibfnamefont
  {Z.}~\bibnamefont {Yang}}, \ and\ \bibinfo {author} {\bibfnamefont
  {P.}~\bibnamefont {Sheng}},\ }\href@noop {} {\bibfield  {journal} {\bibinfo
  {journal} {Appl. Phys. Lett.}\ }\textbf {\bibinfo {volume} {107}},\ \bibinfo
  {pages} {104104} (\bibinfo {year} {2015})}\BibitemShut {NoStop}%
\bibitem [{\citenamefont {Piper}\ \emph {et~al.}(2014)\citenamefont {Piper},
  \citenamefont {Liu},\ and\ \citenamefont {Fan}}]{piper2014}%
  \BibitemOpen
  \bibfield  {author} {\bibinfo {author} {\bibfnamefont {J.~R.}\ \bibnamefont
  {Piper}}, \bibinfo {author} {\bibfnamefont {V.}~\bibnamefont {Liu}}, \ and\
  \bibinfo {author} {\bibfnamefont {S.}~\bibnamefont {Fan}},\ }\href@noop {}
  {\bibfield  {journal} {\bibinfo  {journal} {Appl. Phys. Lett.}\ }\textbf
  {\bibinfo {volume} {104}},\ \bibinfo {pages} {251110} (\bibinfo {year}
  {2014})}\BibitemShut {NoStop}%
\bibitem [{\citenamefont {Merkel}\ \emph {et~al.}(2015)\citenamefont {Merkel},
  \citenamefont {Theocharis}, \citenamefont {Richoux}, \citenamefont
  {Romero-Garc{\'\i}a},\ and\ \citenamefont {Pagneux}}]{merkel2015}%
  \BibitemOpen
  \bibfield  {author} {\bibinfo {author} {\bibfnamefont {A.}~\bibnamefont
  {Merkel}}, \bibinfo {author} {\bibfnamefont {G.}~\bibnamefont {Theocharis}},
  \bibinfo {author} {\bibfnamefont {O.}~\bibnamefont {Richoux}}, \bibinfo
  {author} {\bibfnamefont {V.}~\bibnamefont {Romero-Garc{\'\i}a}}, \ and\
  \bibinfo {author} {\bibfnamefont {V.}~\bibnamefont {Pagneux}},\ }\href@noop
  {} {\bibfield  {journal} {\bibinfo  {journal} {Appl. Phys. Lett.}\ }\textbf
  {\bibinfo {volume} {107}},\ \bibinfo {pages} {244102} (\bibinfo {year}
  {2015})}\BibitemShut {NoStop}%
\bibitem [{\citenamefont {Stinson}(1991)}]{stinson1991}%
  \BibitemOpen
  \bibfield  {author} {\bibinfo {author} {\bibfnamefont {M.~R.}\ \bibnamefont
  {Stinson}},\ }\href@noop {} {\bibfield  {journal} {\bibinfo  {journal} {J.
  Acoust. Soc. Am.}\ }\textbf {\bibinfo {volume} {89}},\ \bibinfo {pages} {550}
  (\bibinfo {year} {1991})}\BibitemShut {NoStop}%
\bibitem [{\citenamefont {Theocharis}\ \emph {et~al.}(2014)\citenamefont
  {Theocharis}, \citenamefont {Richoux}, \citenamefont {Garc{\'\i}a},
  \citenamefont {Merkel},\ and\ \citenamefont {Tournat}}]{theocharis2014}%
  \BibitemOpen
  \bibfield  {author} {\bibinfo {author} {\bibfnamefont {G.}~\bibnamefont
  {Theocharis}}, \bibinfo {author} {\bibfnamefont {O.}~\bibnamefont {Richoux}},
  \bibinfo {author} {\bibfnamefont {V.~R.}\ \bibnamefont {Garc{\'\i}a}},
  \bibinfo {author} {\bibfnamefont {A.}~\bibnamefont {Merkel}}, \ and\ \bibinfo
  {author} {\bibfnamefont {V.}~\bibnamefont {Tournat}},\ }\href@noop {}
  {\bibfield  {journal} {\bibinfo  {journal} {New J. Phys.}\ }\textbf {\bibinfo
  {volume} {16}},\ \bibinfo {pages} {093017} (\bibinfo {year}
  {2014})}\BibitemShut {NoStop}%
\bibitem [{\citenamefont {Pagneux}(2013)}]{pagneux2013}%
  \BibitemOpen
  \bibfield  {author} {\bibinfo {author} {\bibfnamefont {V.}~\bibnamefont
  {Pagneux}},\ }in\ \href@noop {} {\emph {\bibinfo {booktitle} {Dynamic
  Localization Phenomena in Elasticity, Acoustics and Electromagnetism}}}\
  (\bibinfo  {publisher} {Springer},\ \bibinfo {year} {2013})\ pp.\ \bibinfo
  {pages} {181--223}\BibitemShut {NoStop}%
\bibitem [{\citenamefont {Powell}(1978)}]{powell1978}%
  \BibitemOpen
  \bibfield  {author} {\bibinfo {author} {\bibfnamefont {M.~J.}\ \bibnamefont
  {Powell}},\ }in\ \href@noop {} {\emph {\bibinfo {booktitle} {Numerical
  analysis}}}\ (\bibinfo  {publisher} {Springer},\ \bibinfo {year} {1978})\
  pp.\ \bibinfo {pages} {144--157}\BibitemShut {NoStop}%
\bibitem [{\citenamefont {Groby}\ \emph {et~al.}(2010)\citenamefont {Groby},
  \citenamefont {Ogam}, \citenamefont {De~Ryck}, \citenamefont {Sebaa},\ and\
  \citenamefont {Lauriks}}]{groby2010}%
  \BibitemOpen
  \bibfield  {author} {\bibinfo {author} {\bibfnamefont {J.-P.}\ \bibnamefont
  {Groby}}, \bibinfo {author} {\bibfnamefont {E.}~\bibnamefont {Ogam}},
  \bibinfo {author} {\bibfnamefont {L.}~\bibnamefont {De~Ryck}}, \bibinfo
  {author} {\bibfnamefont {N.}~\bibnamefont {Sebaa}}, \ and\ \bibinfo {author}
  {\bibfnamefont {W.}~\bibnamefont {Lauriks}},\ }\href@noop {} {\bibfield
  {journal} {\bibinfo  {journal} {The Journal of the Acoustical Society of
  America}\ }\textbf {\bibinfo {volume} {127}},\ \bibinfo {pages} {764}
  (\bibinfo {year} {2010})}\BibitemShut {NoStop}%
\bibitem [{\citenamefont {Redon}\ \emph {et~al.}(2011)\citenamefont {Redon},
  \citenamefont {{Bonnet-Ben~Dhia}}, \citenamefont {Mercier},\ and\
  \citenamefont {Sari}}]{redon2011}%
  \BibitemOpen
  \bibfield  {author} {\bibinfo {author} {\bibfnamefont {E.}~\bibnamefont
  {Redon}}, \bibinfo {author} {\bibfnamefont {A.-S.}\ \bibnamefont
  {{Bonnet-Ben~Dhia}}}, \bibinfo {author} {\bibfnamefont {J.-F.}\ \bibnamefont
  {Mercier}}, \ and\ \bibinfo {author} {\bibfnamefont {S.~P.}\ \bibnamefont
  {Sari}},\ }\href@noop {} {\bibfield  {journal} {\bibinfo  {journal} {Int. J.
  Numer. Methods Eng.}\ }\textbf {\bibinfo {volume} {86}},\ \bibinfo {pages}
  {1360} (\bibinfo {year} {2011})}\BibitemShut {NoStop}%
\bibitem [{\citenamefont {Kergomard}\ and\ \citenamefont
  {Garcia}(1987)}]{kergomard1987}%
  \BibitemOpen
  \bibfield  {author} {\bibinfo {author} {\bibfnamefont {J.}~\bibnamefont
  {Kergomard}}\ and\ \bibinfo {author} {\bibfnamefont {A.}~\bibnamefont
  {Garcia}},\ }\href@noop {} {\bibfield  {journal} {\bibinfo  {journal} {J.
  Sound Vib.}\ }\textbf {\bibinfo {volume} {114}},\ \bibinfo {pages} {465}
  (\bibinfo {year} {1987})}\BibitemShut {NoStop}%
\bibitem [{\citenamefont {Dubos}\ \emph {et~al.}(1999)\citenamefont {Dubos},
  \citenamefont {Kergomard}, \citenamefont {Khettabi}, \citenamefont {Dalmont},
  \citenamefont {Keefe},\ and\ \citenamefont {Nederveen}}]{dubos1999}%
  \BibitemOpen
  \bibfield  {author} {\bibinfo {author} {\bibfnamefont {V.}~\bibnamefont
  {Dubos}}, \bibinfo {author} {\bibfnamefont {J.}~\bibnamefont {Kergomard}},
  \bibinfo {author} {\bibfnamefont {A.}~\bibnamefont {Khettabi}}, \bibinfo
  {author} {\bibfnamefont {J.-P.}\ \bibnamefont {Dalmont}}, \bibinfo {author}
  {\bibfnamefont {D.}~\bibnamefont {Keefe}}, \ and\ \bibinfo {author}
  {\bibfnamefont {C.}~\bibnamefont {Nederveen}},\ }\href@noop {} {\bibfield
  {journal} {\bibinfo  {journal} {Acta Acustica united with Acustica}\ }\textbf
  {\bibinfo {volume} {85}},\ \bibinfo {pages} {153} (\bibinfo {year}
  {1999})}\BibitemShut {NoStop}%
\bibitem [{\citenamefont {Mechel}(2008)}]{mechel2013}%
  \BibitemOpen
  \bibfield  {author} {\bibinfo {author} {\bibfnamefont {F.~P.}\ \bibnamefont
  {Mechel}},\ }\href@noop {} {\emph {\bibinfo {title} {Formulas of acoustics,
  2nd ed.}}},\ edited by\ \bibinfo {editor} {\bibfnamefont {H.}~\bibnamefont
  {Springer-Verlag}, \bibfnamefont {Berlin}}\ (\bibinfo  {publisher} {Springer
  Science \& Business Media},\ \bibinfo {year} {2008})\ pp.\ \bibinfo {pages}
  {316--327}\BibitemShut {NoStop}%
\end{thebibliography}%

\end{document}